\magnification=\magstep1
\input aua.mac
\MAINTITLE{Maximum-Likelihood Cluster Recontruction}
\AUTHOR{Matthias Bartelmann\at{1,2}, Ramesh Narayan\at{2}, Stella
Seitz\at{1} and Peter Schneider\at{1}}
\INSTITUTE{\at{1}Max-Planck-Institut f\"ur Astrophysik,
Karl-Schwarzschild-Stra{\ss}e 1, D--85748 Garching, Germany;
\at{2}Harvard-Smithsonian Center for Astrophysics, 60 Garden Street,
Cambridge MA 02138, USA}
\ABSTRACT{We present a novel method to recontruct the mass
distribution of galaxy clusters from their gravitational lens effect
on background galaxies. The method is based on a least-$\chi^2$ fit of
the two-dimensional gravitational cluster potential. The method
combines information from shear and magnification by the cluster lens
and is designed to easily incorporate possible additional
information. We describe the technique and demonstrate its feasibility
with simulated data. Both the cluster morphology and the total cluster
mass are well reproduced.}
\titlea{Introduction}
It is of considerable interest for cosmology and the theory of
structure formation in the Universe to learn about the amount and the
spatial distribution of mass in galaxy clusters. While other methods
to determine cluster masses depend on restrictive assumptions about
the symmetry and the equilibrium of the clusters, gravitational
lensing is sensitive to the entire gravitating mass of the lens
regardless of its composition and its dynamical state. The drawback of
lensing-based methods is that they determine the projected mass, thus
giving rise to ambiguities from projection effects, but these should
be compared with the uncertainties of X-ray and dynamical mass
estimates.

Attempts to determine cluster mass distributions from lensing date
back to the suggestion by Webster (1985) that clusters might act as
efficient lenses on background populations of extended objects, and to
the detection of a coherent shear pattern in the field of the cluster
Abell 1689 by Tyson, Valdes, \& Wenk (1990). Kochanek (1990) and
Miralda-Escud\'e (1991) discussed how parameterized cluster mass
distributions could be constrained from observations of weak lensing
in cluster fields. A technique for parameter-free cluster inversion
was pioneered by Kaiser \& Squires (1993), who found that the surface
mass density of galaxy clusters can be derived in terms of a
convolution of the observed shear pattern with a kernel describing the
shear of a point mass. The feasibility of this and related approaches
was then demonstrated by a number of authors (e.g. Bonnet, Mellier, \&
Fort 1994; Fahlman et al. 1994; Smail et al. 1995; Tyson \& Fischer
1995), and Kaiser, Squires, \& Broadhurst (1995a) and Bonnet \&
Mellier (1995) described techniques to accurately determine galaxy
shapes from observed cluster fields.

The cluster inversion technique as proposed by Kaiser \& Squires
(1993) is designed for weak cluster lenses, and it causes unwanted
boundary effects in small or irregularly shaped cluster
fields. Schneider \& Seitz (1995), Seitz \& Schneider (1995a), and
Kaiser (1995) extended the method into the nonlinear regime, and
Schneider (1995), Kaiser et al. (1995b), Bartelmann (1995), and Seitz
\& Schneider (1996) described how the boundary effects can be removed.

All cluster reconstruction techniques based on image distortions alone
can determine the cluster mass distribution only up to a one-parameter
family of linear transformations. This so-called mass-sheet degeneracy
was originally found by Gorenstein, Falco, \& Shapiro (1988) and
discussed in the context of cluster lensing in the weak limit by
Kaiser \& Squires (1993) and generally by Schneider \& Seitz
(1995). The degeneracy can be broken if information on the image
magnification is employed, as suggested by Broadhurst, Taylor, \&
Peacock (1995) and Bartelmann \& Narayan (1995).

Unambiguous determinations of cluster masses from weak lensing require
the redshift distribution of background sources to be known. Apart
from direct spectroscopy, which is hampered by the faintness of the
background sources, lensing itself provides means to find faint-galaxy
redshifts. Kneib et al. (1994) estimated arclet redshifts in the field
of the cluster Abell 370, and Smail, Ellis, \& Fitchett (1994)
discussed how faint-galaxy redshifts can be constrained from weak
lensing by a number of clusters. Bartelmann \& Narayan (1995) proposed
an algorithm to simultaneously derive cluster mass distributions and
the redshift distribution of faint galaxies.

We propose here a novel cluster-reconstruction technique which is
local, thus avoiding boundary effects by design, and which combines
information on galaxy distortions and magnifications, thus breaking
the mass-sheet degeneracy. The key idea is to reconstruct the
two-dimensional Newtonian potential of the cluster lenses with a
least-$\chi^2$ approach. The least-$\chi^2$ method lends itself to
straightforwardly including additional information into the cluster
reconstruction process. We aim at the gravitational potential rather
than at the surface-mass density because it is the physical object
underlying both lensing distortion and magnification, and because it
should simplify comparisons of determinations of the cluster mass
distribution from gravitational lensing with those from other means
like, e.g., interpretations of the X-ray emission.

This paper outlines the method and demonstrates its feasibility using
simulated data. Extensions and a detailed discussion are postponed to
a later, more technical paper (Seitz et al., in preparation). Section
2 describes the technique. In section 3, we present results from
numerical simulations, and we summarize the paper in section 4.
\titlea{Outline of the method}
Gravitational lensing magnifies and distorts images of extended
background sources. The local properties of the lens mapping are
described by the Jacobian
$$
  {\cal A} = {\partial\vec y\over\partial\vec x} =
  \pmatrix{1-\kappa-\gamma_1 & -\gamma_2 \cr
           -\gamma_2 & 1-\kappa+\gamma_1 \cr}\;,
\eqno(1)$$
where $\vec x$ and $\vec y$ are dimensionless two-dimensional position
vectors in the lens- and source planes, respectively. The convergence
$\kappa$ is the scaled surface mass density of the lens, and
$\gamma_{1,2}$ are the components of the shear. They are combinations
of second derivatives of the (dimensionless) two-dimensional Newtonian
potential of the lens,
$$
  \kappa = {1\over2}\left(\psi_{,11}+\psi_{,22}\right)\;,\quad
  \gamma_1 = {1\over2}\left(\psi_{,11}-\psi_{,22}\right)\;,\quad
  \gamma_2 = \psi_{,12}\;.
\eqno(2)$$
For general reference on lensing, see Schneider, Ehlers, \& Falco
(1992) or Blandford \& Narayan (1992). Image distortions, which can
be quantified by, e.g., the quadrupole tensor of the
surface-brightness distribution, measure the two-component quantity
$$
  g_i = {\gamma_i\over1-\kappa}
\eqno(3)$$
(Miralda-Escud\'e 1991; Schneider 1995; Kaiser 1995). Since the
ellipticity of an image is independent of its size, the ellipticity is
unchanged if the Jacobian is multiplied by a factor $\lambda\ne0$,
which corresponds to transforming $\kappa$ and $\gamma_i$ according to
$$
  (1-\kappa)\to\lambda(1-\kappa)\;,\quad\gamma_i\to\lambda\gamma_i\;.
\eqno(4)$$
The lens strength $g_i$ above is manifestly invariant under this
transformation, which reflects the mass-sheet degeneracy mentioned in
the introduction. If the cluster has a critical curve, an ambiguity
arises in the $g_i$ because of the parity change upon crossing the
critical curve. An unambiguous measure of the ellipticity is then
provided by the distortion $\delta_i$,
$$
  \delta_i = 
  {2\gamma_i\,(1-\kappa)\over(1-\kappa)^2+\gamma_1^2+\gamma_2^2}
$$
(Schneider \& Seitz 1995; Miralda-Escud\'e 1991).

The magnification $\mu$ is given by
$$
  \mu = {\det}^{-1}{\cal A} =
  \left[(1-\kappa)^2-\gamma_1^2-\gamma_2^2\right]^{-1}\;,
\eqno(5)$$
which scales with $\lambda^{-2}$ under the transformation
(4). Measurements of the magnification can therefore be used to break
the mass-sheet degeneracy, as Broadhurst et al. (1995) recognized. The
magnification is accessible on a statistical basis by comparing the
sizes of galaxies in cluster fields with those of galaxies of equal
surface brightness in empty fields or by the change in number density
of galaxies (Bartelmann \& Narayan 1995; Broadhurst et al. 1995).

We assume in the following that the sizes and shapes of galaxies in a
cluster field and in an empty control field have been determined. The
data are averaged over suitable regions of the cluster field to reduce
the noise. These regions have to be small enough so that the
properties of the lens can be assumed to be constant across each
region, and large enough such that they contain a sufficient number of
galaxies. Since the number of background galaxies is of the order of
$10^5$ per square degree at $B\approx27$ (e.g. Tyson 1994), this
requirement is easily met. We do the smoothing with a Gaussian
filter. The smoothing length is spatially constant and is varied later
to minimize the $\chi^2$ obtained in the fit.

For simplicity, we assume in the following that we already have the
smoothed data on a regular grid covering the cluster field, i.e., that
we have obtained smoothed measurements $g_i(k,l)$ and $\mu(k,l)$ of
the two components of the lens strength and the magnification in each
cell $(k,l)$ of the grid. For simplicity of notation, we use the
inverse magnification $r(k,l)\equiv\mu^{-1}(k,l)$ rather than the
magnification.

Of course, local estimates of the surface-mass density can be obtained
directly once $g$ and $r$ are given at each grid cell (Broadhurst
1996). However, such an approach has the major disadvantage that it
neglects the fact that convergence and shear are intrinsically related
fields. This relation provides important additional information, and
thus considerably reduces the noise of the reconstruction. In our
approach, the one physical field $\psi$ underlying all local lensing
effects is reconstructed, and it is determined such as to optimize the
global agreement of $\psi$ with the data.

We now want to determine least-$\chi^2$ fits to the values of the
potential $\psi(k,l)$ in the cells $(k,l)$ of the data grid. To do so,
we replace the second partial derivatives of the potential in eq. [2]
by their second-order finite-difference approximations. The resulting
expressions can then be used to form estimators $\hat g_i(k,l)$ and
$\hat r(k,l)$ in terms of $\psi$ for the lens strength $g_i(k,l)$ and
the inverse magnification $r(k,l)$ at the grid cell $(k,l)$. The
appropriate $\chi^2$ function then reads
$$
  \chi^2 = \sum_{k,l}\left\{
  {1\over\sigma_g^2(k,l)}\left[g_i(k,l)-\hat g_i(k,l)\right]^2 +
  {1\over\sigma_r^2(k,l)}\left[r(k,l)-\hat r(k,l)\right]^2\right\}\;,
\eqno(6)$$
where summation over the two components of $g_i$ is implied. Of
course, $\chi^2$ depends on the values $\psi(m,n)$ of the potential at
the grid points $(m,n)$. The variances $\sigma_g(k,l)$ and
$\sigma_r(k,l)$ of $g$ and $r$ can be estimated directly from the
data. Since the derivatives
$$
  {\partial\chi^2\over\partial\psi(m,n)}
$$
are straightforwardly determined, $\chi^2$ can then be minimized
numerically using a conjugate-gradient algorithm (e.g. Press et
al. 1992, section 10.6).

It is implicitly assumed in eq. [6] that $g$ and $r$ can be determined
independently, which is the case in our simulation. In practice,
however, $g$ and $r$ can be correlated especially for faint images. In
that case, one would determine the local averages of $g$ and $r$ by
giving lesser weight to fainter images, and modify the definition of
$\chi^2$ such as to account for a possible correlation.

The grid for the potential coincides with the grid of the smoothed
data within the observed field, but it has to be extended by one
further row or column of grid cells along each side of the field to
allow the partial derivatives to be determined at any grid point where
data are given. We start the potential fit with $\psi(m,n)=0$ for all
$(m,n)$. Since only second derivatives of the potential enter into the
algorithm, we are free to add an arbitrary constant and a spatially
constant gradient to the potential. We choose to fix the potential to
$\psi=0$ at three corners of the field. The result is a least-$\chi^2$
fit of the potential values $\psi(m,n)$ on the grid, from which
$\kappa$ and thus the surface-mass density can be derived via
eq. [2]. In short, the $\chi^2$ minimization determines the potential
such as to optimize the agreement between the measured ellipticities
and sizes of the lensed galaxies and the shear and the convergence
expected from the potential.
\titlea{Numerical simulations}
For testing the method, we use a numerically simulated galaxy cluster
from the sample described in detail by Bartelmann, Steinmetz, \& Weiss
(1995). Briefly, the cluster was simulated within a COBE-normalized
CDM cosmological simulation, taking the tidal forces of the
surrounding structures into account. The cluster model we select is at
redshift $z_{\rm d}=0.16$. It is resolved into $\sim17.000$ particles
whose line-of-sight velocity dispersion is $\sim1200\;$km$\,$s$^{-1}$.

Background galaxies are assumed to be at a constant redshift of
$z_{\rm s}=1$. They are assigned random intrinsic ellipticity
components $(\varepsilon_1,\varepsilon_2)$ drawn from the probability
distribution
$$
  p_{\rm e}(\varepsilon_1,\varepsilon_2) = 
  {\exp\left(-|\varepsilon|^2/\sigma_\varepsilon^2\right)\over
  \pi\sigma_\varepsilon^2
  \left[1-\exp\left(-1/\sigma_\varepsilon^2\right)\right]}\;,
\eqno(7)$$
where $|\varepsilon|^2\equiv\varepsilon_1^2+\varepsilon_2^2$. For
elliptical images, the modulus of $\varepsilon$ is given by
$$
  |\varepsilon| = {a-b\over a+b}\;,
$$
where $a$ and $b$ are the major and minor axes of the ellipse,
respectively. We choose $\sigma_\varepsilon=0.15$
(cf. Miralda-Escud\'e 1991; Tyson \& Seitzer 1988; Brainerd,
Blandford, \& Smail 1995). Further galaxy properties are chosen
according to the galaxy model motivated and described by Bartelmann \&
Narayan (1995). They include luminosity and surface brightness, which
allow to assign an intrinsic size to the sources. The logarithmic
variance of the intrinsic galaxy radii is $\Delta\ln(R)\approx0.5$, in
good agreement with {\it HST} measurements (Kaiser 1994, private
communication).

We simulate the lensing effect of the cluster on the sources, assuming
a source density of 70 galaxies per square arc minute. In addition, we
simulate an independent empty galaxy field to calibrate the intrinsic
sizes of the sources needed for our algorithm. The field side length
is $5'$, and the field is covered by a grid of $10\times10$ cells (or
$12\times12$ cells for the potential). The simulated galaxy data in
the lensed and the unlensed fields are then analyzed with the
least-$\chi^2$ potential reconstruction method described
previously. We vary the smoothing length such as to minimize the
$\chi^2$ obtained by the potential fit. Various results are presented
in figures 1 and 2. The $\chi^2$ per degree of freedom in this case is
$1.08$, with about equal contributions from the distortion and the
size information.

\begfig{fig1.tps}
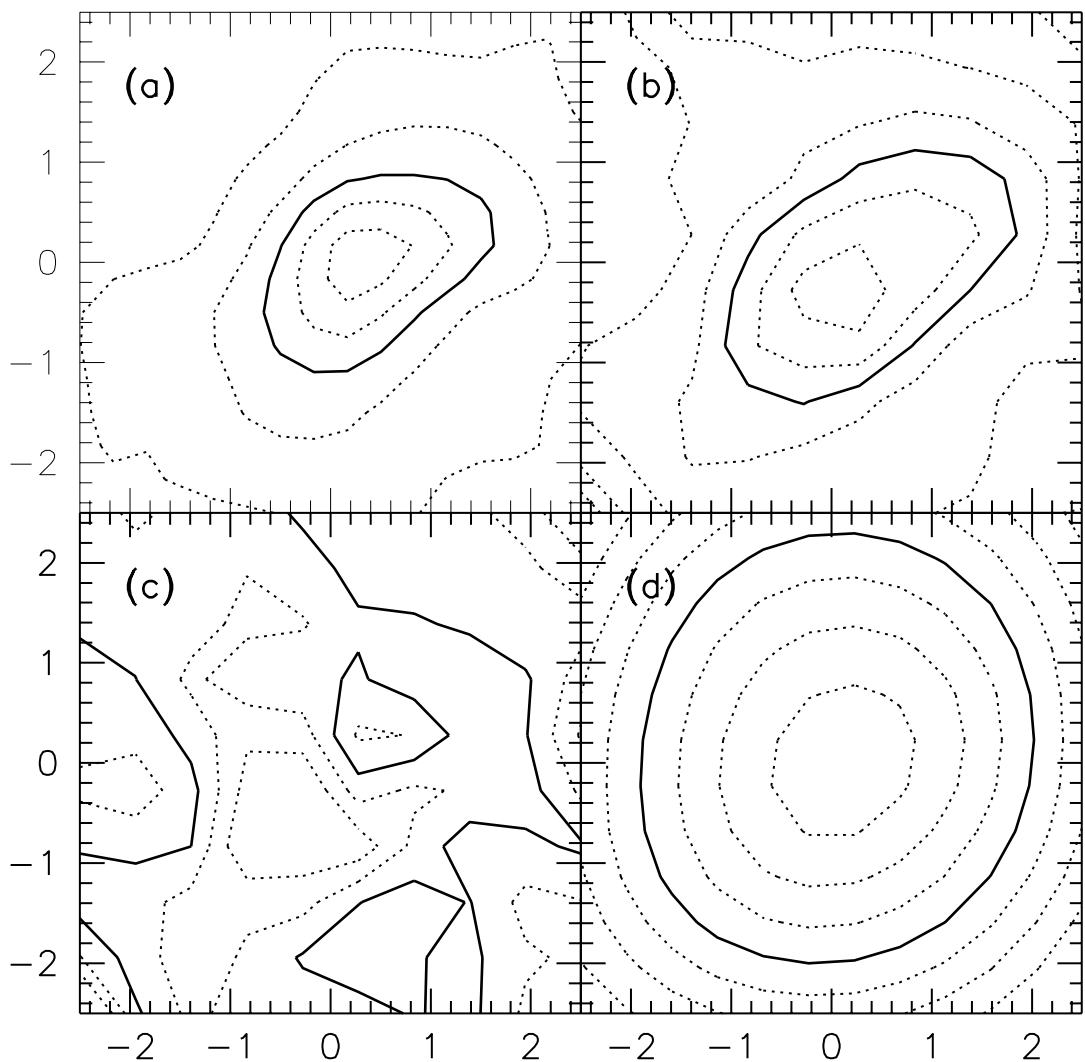
\figure{1}{Four contour plots showing the original cluster model in
panel (a), the reconstruction in panel (b), the difference between the
two in panel (c), and the dimensionless two-dimensional potential in
panel (d). Contours in panels (a) and (b) are spaced by $0.1$ and the
heavy contour follows $\kappa=0.5$. In panel (c), contours are spaced
by $0.05$ and the heavy contour follows $\Delta\kappa=0$. The
potential is kept fixed at $\psi=0$ at three corners. The heavy line
in panel (d) follows the arbitrary contour $\psi=-5$, and the contours
are spaced by $1.5$. The side length of the fields is $5'$.}
\endfig

Figure 1 shows contour plots of the original cluster model in panel
(a), the reconstruction in panel (b), the difference between these two
in panel (c), and the potential in panel (d). A comparison between the
upper panels shows that the cluster is very well reproduced except
that the involved smoothing of the data tends to broaden the central
mass peak. The agreement is quantified in panel (c), which shows that
the residual is small and distributed fairly homogeneously across the
field. In particular, there are no systematic deviations towards the
field boundary.

\begfig{fig2.tps}
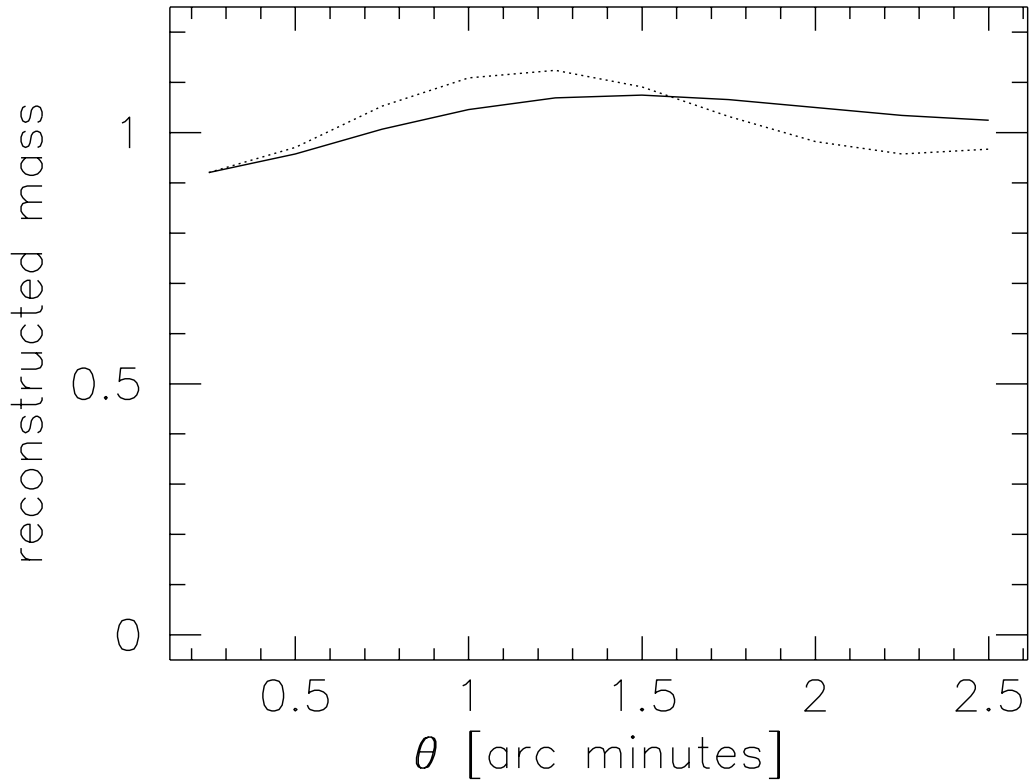
\figure{2}{The reconstructed mass fraction within circles centered on
the cluster center. The solid line shows the fraction of the
cumulative mass within radius $\theta$, the dotted curve the mass
fraction within annuli. $\theta$ is given in arc minutes.}
\endfig

Figure 2 displays two curves. The solid curve shows the total mass
within radius $\theta$ from the center of the cluster relative to the
input mass, i.e., the cumulative mass fraction within circles as a
function of the circle radius. The dotted curve shows the mass
fraction in annuli with outer radius $\theta$ and width
$0\farcm25$. The solid curve starts at $\sim0.92$, rises to
$\sim1.07$, and falls towards $\sim1$ at the field boundary. This
shows that the total cluster mass enclosed by circles is reproduced to
$\sim7\%$ at any radius, and to higher accuracy at the field
boundary. The mass fraction in annuli shows a somewhat larger
fluctuation of $\sim\pm10\%$. Both curves underestimate the mass in
the cluster center, because the smoothing involved in the preparation
of the input data tends to broaden the central mass peak, thereby
shifting mass to larger radii. This local overestimate at intermediate
radii is compensated by a slight underestimate at larger radii such as
to reproduce the total mass, which is fixed by the overall
magnification effect of the cluster.
\titlea{Summary}
We have suggested a novel method to reconstruct cluster mass
distributions from their gravitational lens effect on a population of
background sources. Both distortion and magnification effects are
included. The method is based on a least-$\chi^2$ fit of the
two-dimensional Newtonian potential of the lensing
cluster. Reconstructing the potential is advantageous because it is
the physical quantity underlying both the distortion and the
magnification. The least-$\chi^2$ approach has the advantages that it
provides a local reconstruction technique, which renders it
insensitive to boundary effects, and that it can be almost arbitrarily
extended to encorporate additional information. The combination of
shear and magnification effects breaks the mass-sheet degeneracy
inherent in all reconstruction algorithms based on ellipticity
information alone.

Application of our method requires to measure not only the
ellipticities of background galaxies, but also their intrinsic and
their magnified sizes. Although the intrinsic dispersion of the sizes
is large, the large number density of background galaxies supplies a
sufficiently large data base to statistically extract reliable
information on the magnification. The intrinsic sizes of the
background galaxies are to be obtained from observations in empty
fields.

We have demonstrated using numerical simulations that the method
accurately reproduces the shape and the total mass in the lensing
cluster. The lensing cluster was numerically simulated within a CDM
scenario, taking the tidal field of the surrounding matter
distribution into account. The numerically simulated population of
background galaxies was adapted to observations. It is designed to
reproduce the observed intrinsic scatter in galaxy sizes and
ellipticities and the number density of these objects.

Various modifications and refinements are possible and can be included
into the proposed technique. Among them are the effect of the lensing
magnification on the number density of background galaxies (originally
discussed by Broadhurst et al. 1995), and regularized
maximum-likelihood techniques to avoid the (to some degree arbitrary)
smoothing of the data and to ensure that the fitted data points are
independent. Such modifications and extensions will be discussed in
a following paper (Seitz, Schneider, Bartelmann, \& Narayan, in
preparation).
\acknow{We thank Bill Press for valuable comments and Matthias
Steinmetz for the numerical simulation of the cluster model. This work
was partially supported by the Sonderforschungsbereich SFB 375-95 of
the Deutsche Forschungsgemeinschaft (MB, SS, PS), and by NSF grants
AST 9423209 and PHY 94-07194 (RN).}
\begref{References}
\ref Bartelmann, M. 1995, A\&A, 303, 643
\ref Bartelmann, M., \& Narayan, R. 1995, ApJ, 451, 60
\ref Bartelmann, M., Steinmetz, M., \& Weiss, A. 1995, A\&A,
297, 1
\ref Blandford, R. D., \& Narayan, R. 1992, ARA\&A, 30, 311
\ref Bonnet, H., \& Mellier, Y. 1995, A\&A, 303, 331
\ref Bonnet, H., Mellier, Y., \& Fort, B. 1994, ApJ, 427, L83
\ref Brainerd, T. G., Blandford, R. D., \& Smail, I. 1995,
preprint
\ref Broadhurst, T. J. 1996, ApJL, submitted
\ref Broadhurst, T. J., Taylor, A. N., \& Peacock, J. A. 1995,
ApJ, 438, 49
\ref Fahlman, G., Kaiser, N., Squires, G., \& Woods, D. 1994,
ApJ, 437, 56
\ref Gorenstein, M. V., Falco, E. E., \& Shapiro, I. I. 1988,
ApJ, 327, 693
\ref Kaiser, N. 1995, ApJ, 493, L1
\ref Kaiser, N., \& Squires, G. 1993, ApJ, 404, 441
\ref Kaiser, N., Squires, G., \& Broadhurst, T. J. 1995a, ApJ,
449, 460 
\ref Kaiser, N., Squires, G., Fahlman, G., Woods, D., \&
Broadhurst, T. 1995b, in press
\ref Kneib, J.-P., Mathez, G., Fort, B., Mellier, Y., Soucail,
G., \& Longaretti, P.-Y. 1994, A\&A, 286, 701
\ref Kochanek, C. S. 1990, MNRAS, 247, 135
\ref Miralda-Escud\'e, J. 1991, ApJ, 370, 1
\ref Press, W. H., Teukolsky, S. A., Vetterling, W. T., \&
Flannery, B. P. 1992, Numerical Recipes (Cambridge: University Press)
\ref Schneider, P., 1995, A\&A, 302, 639
\ref Schneider, P., \& Seitz, C. 1995, A\&A, 294, 411
\ref Schneider, P., Ehlers, J., \& Falco, E. E. 1992,
Gravitational Lenses (Heidelberg: Springer)
\ref Seitz, C., \& Schneider, P. 1995a, A\&A, 297, 287
\ref Seitz, S., \& Schneider, P. 1996, A\&A, 305, 383
\ref Smail, I., Ellis, R. S., \& Fitchett, M. J. 1994, MNRAS,
270, 245
\ref Smail, I., Ellis, R. S., Fitchett, M. J., \& Edge,
A. C. 1995, MNRAS, 273, 277
\ref Tyson, J. A. 1994, in: Cosmology and Large-Scale
Structure, Proc. Les Houches Summer School 1993, ed. R. Schaeffer et
al.
\ref Tyson, J. A., \& Seitzer, P. 1988, ApJ, 335, 552
\ref Tyson, J. A., \& Fischer, P. 1995, ApJ, 446, L55
\ref Tyson, J. A., Valdes, F., \& Wenk, R. A. 1990, ApJ, 349,
L1
\ref Webster, R. L. 1985, MNRAS, 213, 871
\endref
\end